# Gate-tunable and high responsivity graphene phototransistors on undoped semiconductor substrates

Biddut K. Sarker,[1,2] Isaac Childres,[1,2] Edward Cazalas,[3] Igor Jovanovic,[3] and Yong P. Chen[1,2,4,*]

[1]Department of Physics and Astronomy, Purdue University, West Lafayette, Indiana 47907, USA
[2]Birck Nanotechnology Center, Purdue University, West Lafayette, Indiana 47907, USA
[3]Department of Mechanical and Nuclear Engineering, The Pennsylvania State University, University Park, Pennsylvania 16802, USA
[4]School of Electrical and Computer Engineering, Purdue University, West Lafayette, Indiana 47907, USA
Corresponding Author. Email: yongchen@purdue.edu

Abstract

Due to its high charge carrier mobility, broadband light absorption, and ultrafast carrier dynamics, graphene is a promising material for the development of high-performance photodetectors. Graphene-based photodetectors have been demonstrated to date using monolayer graphene operating in conjunction with either metals or semiconductors. Most graphene devices are fabricated on doped Si substrates with $SiO_2$ dielectric used for back gating. Here, we demonstrate photodetection in graphene field effect phototransistors fabricated on *undoped* semiconductor (SiC) substrates. The photodetection mechanism relies on the high sensitivity of the graphene conductivity to the local change of the electric field that can result from the photo-excited charge carriers produced in the back-gated semiconductor substrate. We also modeled the device and simulated its operation using the finite element method to validate the existence of the field induced photoresponse mechanism and study its properties. Our graphene phototransistor possesses a room-temperature photoresponsivity as high as ~ 7.4 A/W, higher than the required photoresponsivity (1 A/W) in most practical applications. The light power-dependent photocurrent and photoresponsivity can be tuned by the source-drain bias voltage and back-gate voltage. Graphene phototransistors based on this simple and generic architecture can be fabricated by depositing graphene on a variety of undoped substrates, and are attractive for many applications in which photodetection or radiation detection is sought.

Graphene, a single layer of carbon atoms in a honeycomb lattice, is a fascinating new material with a potential for use in a variety of applications, including the next-generation electronic and optoelectronic devices.[1-5] In particular, graphene-based photodetectors have attracted significant attention due to their relatively wide absorption spectrum, high carrier mobility, low cost, and feasibility of their integration into flexible and transparent devices.[6-14]

Over the last several years, a variety of graphene-based photodetectors have been reported,[6-24] which can be classified into two primary types (type I and type II). In type I graphene-based photodetectors, graphene is used to both generate and transport photoexcited carriers. The mechanisms of photodetection identified in these photodetectors include the photovoltaic effect,[6,7,15,18] photothermoelectric effect,[8,16,17,23,24] and bolometric effect.[18,21] Regardless of the photodetection mechanism, photoresponse of the type I graphene-based photodetectors is very fast (on the order of GHz). However, the photoresponsivity (photocurrent per unit power of incident light) is relatively low (on the order of mA/W), largely due to the weak light absorption by graphene.[6,7,12,13] Various techniques, such as integrating graphene with photonic nanostructures (e.g. microcavities, waveguides, plasmonic arrays, etc.), have been proposed to increase the photoresponsivity by increasing the light absorption, but the photoresponsivity has not been improved to more than a few tens of mA/W.[11,25-28] Moreover, the fabrication procedures of these photodetectors are relatively complex. Recently, band structure engineering in graphene has enhanced photoresponsivity but only at temperatures below ~ 200 K, while the photoresponsivity at room temperature is still relatively low.[20] In the type II graphene-based photodetectors, a combination of graphene and other photoactive nanomaterials (such as semiconductor quantum dots) are used.[29-31] In this case, photo-carriers are generated in the photoactive materials and then transferred to and transported by graphene, which acts as a conducting channel. Using this approach, the photoresponsivity is increased significantly by the virtue of the much higher photo-absorption by the photoactive nanomaterials.[29,30]

In this work, we demonstate a simple approach to graphene-based photodetection that utilizes a photo-actuated graphene field effect transistor (GFET) on a back-gated *undoped* semiconductor substrate (Fig. 1a). In such a graphene phototransistor, the photodetection mechanism relies on the high sensitivity of the conductivity of graphene to the local change of the electric field, resulting from photo-excited carriers produced in the underlying electrically gated, undoped semiconductor substrate. We validate this field effect based photodetection mechanism with the finite element method (FEM) simulations of the electric and potential field distribution within the GFET for different laser powers (Fig. 1d). We show that the GFETs fabricated on undoped silicon carbide (SiC) substrates exhibit a high photoresponsivity of ~7.4 A/W at room temperature. This is about three orders of magnitude higher than photoresponsivity of type-I photodetectors graphene photodetectors (typically fabricated on doped silicon substrates).[7,18] The photocurrent and photoresponsivity of the GFETs based on this novel architecture can be tuned by the gate voltage and source-drain bias voltage and is dependent on the incident optical power. The response time is reduced significantly with increasing incident optical power. The shortest response time measured in our device is ~ 1 s. The methodology presented here can provide a new and simple pathway for the development of high-responsivity graphene photodetectors (particularly for applications where a high speed of response is not essential).

A typical device architecture of the GFET on an undoped semiconductor substrate (SiC in our case) is shown in Fig. 1a. Our devices were fabricated by transferring mechanically exfoliated single layer graphene onto 416 μm thick SiC substrates, followed by electron beam lithography and deposition of Cr/Au contacts (see Methods and Supplementary Information for details of the fabrication process). An optical image (top view) of a part of a fabricated device is displayed in the inset of Fig. 1b. The presence of single layer graphene is confirmed by Raman spectroscopy (Inset of Fig. 1c, and Supplementary Information). The optoelectronic measurements were performed by illuminating the entire device with a laser beam (wavelength $\lambda$ = 400 nm, corresponding photon energy = 3.1 eV). All measurements were performed at room temperature and atmospheric pressure.

Figure 1b shows the measured drain-source current ($I_{ds}$) as a function of the back-gate voltage ($V_g$) of a representative device without ("dark") and with laser illumination ("light"). Without illumination, the effect of $V_g$ on the dark current ($I_{dark}$) is relatively small. When the device is illuminated, the field effect response is significantly *enhanced*, as demonstrated by a larger gate voltage modulation of the current under laser illumination ($I_{light}$, Fig. 1b), suggesting that the (same) gate voltage now exerts a stronger electric field on graphene. In Fig. 1c, the photocurrent ($I_{photo}$, defined as $I_{light}$ - $I_{dark}$) is displayed as a function of $V_g$. The $I_{photo}$-$V_g$ plot shows that the photocurrent is positive for sufficiently negative $V_g$ but undergoes a sign reversal near $V_g$ ~ 0V and becomes negative for positive $V_g$. Thus, both the polarity and magnitude of photocurrent of our device can be tuned with the gate voltage.

The observed gate voltage-dependent photocurrent of our GFETs can be qualitatively explained by the following mechanism. Under dark condition, the undoped SiC is highly insulating (bandgap ~3 eV)[32] and the applied $V_g$ drops uniformly across the SiC substrate. Due to the relatively large thickness ($d$ ~ 416 μm) of our SiC substrate, the electric field ($E = V_g/d$) experienced by graphene is relatively small, giving rise to a weak field effect. The observed finite small field effect without illumination may arise from the residual conductivity of the SiC due to impurities or trapped charges. When the SiC is illuminated, photo-excited charge carriers are generated in SiC, leading to an increased conductivity. While the SiC becomes more conductive under illumination, the experimentally observed leakage current between the back gate and graphene does not increase notably (the SiC does not form a shorted connection between the backgate and graphene). This can be due to the presence of a native oxide layer that often forms naturally on the SiC surface.[33,34] This native oxide (whose bandgap is much larger than our laser photon energy) remains insulating even under illumination. This could also arise from the spatially non-uniform distribution of the photogenerated carriers in SiC, where parts of the SiC may remain insulating under illumination. The enhanced field effect seen in Fig. 1b suggests that with increasing conductivity of SiC, the electric field at the graphene due to the applied back gate voltage increases. Such a photo-actuated change in the electric field is sensed by the change of graphene conductivity *via* field effect, allowing us to detect the light interacting with SiC.[35] This proposed mechanism is also consistent with the observation of near-zero photocurrent at $V_g$ ~ 0 V, where there is no electric field (and thus no field effect) to modulate the graphene conductivity. The small offset of zero crossing point of photocurrent away from $V_g$ = 0 V (Fig. 1c) may be related to gate hysteresis and trapped charges in the SiC.[36]

To better understand the field effect based photodetection mechanism, we conducted FEM simulations of the electric field and potential distribution within the SiC substrate in the GFET using COMSOL Multiphysics.[35] The results of our simulation are presented in Fig. 1d. The architecture and thickness of the SiC substrate used in the modeled device closely match that of our experimental devices. To qualitatively capture the effect of the native oxide and the part of SiC substrate that remains insulating under illumination, we assumed the conductivity of the top 10 nm portion of our SiC substrate to be unaffected by illumination in our simulation. The laser illumination modifies the electric field within the SiC *via* the change in conductivity within SiC (except the top 10 nm). The conductivity of SiC affected by illumination is calculated for different incident laser power (see Supplementary Information) and is used as an input to the model which simulates the electric field. A representative result showing the simulated electric potential in the SiC for a laser power of 100 µW is displayed in the inset of Fig. 2d. The calculated electric field under graphene (at the SiC/graphene interface) for various laser powers is plotted in Fig. 2d. The model generally shows that for increasing illumination power a greater electric field exists in the vicinity of graphene (Fig. 1d), which results in the modulation of conductivity of graphene. This change of graphene conductivity is used to detect the light incident on the GFET.

To further validate the proposed mechanism of photoresponse in our GFETs, we fabricated and measured two control devices. One is a SiC device fabricated by making contacts on top of a bare SiC substrate (no graphene in the channel), and another is a dummy device in which gold is used as the channel instead of graphene (see Supplementary Information). The SiC device shows a very small photocurrent (on the order of nA, ~ 0.1 % of graphene photocurrent), whereas the dummy device shows no change in current with change in gate-voltage, both in dark and under light illumination. These measurements confirm that the photocurrent of our GFETs does not result from the Schottky contact at the metal/SiC interface or the field effect from the SiC; rather, the photocurrent originates from the modulated charge carriers in the graphene due to graphene field effect. In addition to these control experiments, we also measured the gate leakage current of the GFET, finding that it is small (<1 nA at $V_g = \pm$ 30 V), even with light illumination (see Supplementary Information), much lower than the measured photocurrent, which can reach many tens of µA. This further confirms that photocurrent of our GFET is not the result of collection of charge from the SiC.

We further studied the dependence of photoresponse on the source-drain bias voltage ($V_{ds}$) at different illumination powers. Fig. 2a shows the $I_{ds}$ - $V_{ds}$ characteristics of a typical device without and with illumination for a series of incident laser power $P_{in}$ (varying from 1 to 184 µW) for a representative $V_g = -$ 20 V. We found that all $I_{ds}$ - $V_{ds}$ curves pass through the origin, while the slope (indicating the conductance of graphene) of $I_{ds}$ - $V_{ds}$ curves increase with increasing laser power. Fig. 2b displays an enlarged view of the circled region shown in (a), showing that $I_{light}$ increases with increasing $P_{in}$. Using the data in Fig. 2a, we calculated the photocurrent ($I_{photo}$) and plotted its dependence on $V_{ds}$ in Fig. 3a. For all $P_{in}$, $I_{photo}$ increases linearly with increasing $V_{ds}$, and a large photocurrent ~ 34 µA is observed for $V_{ds} = -$ 0.5 V and $P_{in} = 184$ µW.

One of the most important figures of merit of a photodetector is its photoresponsivity ($R$), defined as the ratio of photocurrent and the incident laser power, $R = I_{photo}/P_{in}$. The plots of $R$ as a function of $V_{ds}$ at different laser powers (Fig. 3b) show that $R$ increases linearly with increasing $V_{ds}$, suggesting that the device is in the linear response regime, and the photoresponsivity can be increased by applying a higher $V_{ds}$. For $V_{ds} = -$ 0.5 V, our device shows a high photoresponsivity of 7.4 A/W, which is more than three orders of magnitude higher than that previously measured in the ("type-I") graphene photodetectors (with a similar or higher $V_{ds}$).[7,18,25-28] Although the photoresponsivity of our device is lower than that observed in the graphene-quantum dot hybrid phototransistors,[29,30] it is still larger than the required photoresponsivity (~1 A/W) for most practical applications.[3,14] Moreover, the simple device architecture and fabrication process required by this architecture may offer significant practical advantages. We also note that the photoresponsivity we report here (photoresponsivity is calculated using the *total* laser power incident on our entire device, not considering only the area of graphene) is likely to be underestimated, since a part of the laser beam incident on the SiC far away from the graphene may not contribute significantly to the observed photoresponse.

We attribute this high photoresponsivity of our device to the unique device architecture, which supports an entirely different photodetection mechanism. Unlike the type-I graphene photodetectors reported to date, [6,7,15,18] in our devices the undoped SiC substrate is employed as a light absorber. In the presence of the back-gate voltage the photoexcited carriers in the SiC modulate the electric field, thus also inducing the charge carriers in graphene *via* a field effect. The highly sensitive field effect of graphene provides an efficient intrinsic amplification mechanism that (indirectly) converts the photon energy into a large electrical signal and hence leads to a high photoresponsivity. The number of modulated charge carriers (electrons or holes) in the graphene per incident photon in our device can reach as high as ~23 at a laser power of 1 µW (see Supplementary Information).

More insight into the photoresponse characteristics of our device can be obtained from the dependence of photocurrent and photoresponsivity on the incident laser power $P_{in}$. As shown in Fig. 3c, at lower $P_{in}$ (for example, below ~15 µW for $V_{ds} = -0.5$ V), the photocurrent increases with increasing $P_{in}$ due to an increase in the modulated charge carriers. However, at higher $P_{in}$, the photocurrent saturates (Fig. 3c), leading to a decrease in the photoresponsivity, as shown in Fig. 3d. One possible reason for this observed photocurrent saturation could be the saturation of graphene field effect itself at large (modulated) charge carrier densities (seen also in Fig. 1b) due to factors such as contact resistance and existence of charge trap states in graphene or at the graphene-SiC interface. The saturation might also result from decreased electric field modulation in the substrate at higher incident optical powers. We found that the decrease of $R$ with increasing $P_{in}$ can be fitted by a power law, $R \propto P_{in}^{\beta}$ with $\beta \sim -0.8$ (inset of Fig. 3d). We note that similar power law relations have been observed in phototransistors based on graphene-MoS$_2$ hybrid with $\beta \sim -0.8$,[37] and based on black phosphorus with $\beta \sim -0.3$,[38] (in the latter work this was attributed to the reduction of photogenerated carriers at the higher power due to the recombination/trap states).[38]

We now turn our attention to the transient photoresponse of our devices. Time-dependent photocurrent for different representative gate voltages were measured as the laser was turned on and off (Fig. 4a). It is found that the sign of photocurrent changes from positive to negative as $V_g$ changes from -20 V to +20 V, and the photocurrent is almost zero for $V_g = 0$ V. Both features are consistent with the field effect measurement shown in the Fig. 1c and confirm the gate tunability of our device's photoresponse. The gate-tunability is important for photodetection since it offers a convenient on-off switching control. In addition to the gate-tunability, our device maintains a long-term stability and a good reproducibility of the photoresponse for a series of repeated laser on/off switching, as shown in Fig. 4b.

We found that the characteristics of time-dependent photocurrent curves vary significantly with increasing laser power (Fig. 4c). For a laser power of 184 µW, the photocurrent to dark-current ratio ($I_{photo}/I_{dark}$) of our device can reach up to 10.3% (inset of Fig. 4d), which is higher than that of other recently reported graphene devices.[39, 40] We calculated photocurrent response time ($\tau$) by fitting the experimental data in Fig. 4c to an exponential function (see Supplementary Information) and plotted $\tau$ as a function of $P_{in}$ in Fig. 4d. We find that the response time for photocurrent rise ($\tau_{rise}$) and fall ($\tau_{fall}$) for each $P_{in}$ are similar. As the laser power increases, the response time decreases and can be fitted with a power law, $\tau \propto P_{in}^{\alpha}$, with $\alpha \sim -0.7$. The shortest response time of our device is ~ 1 s (measured at the highest $P_{in}$ = 184 µW), which is similar to the response time of a graphene-quantum dot hybrid photodetector.[30] The possible reasons for long response time could be due to the electrochemical doping of graphene.[36]

In summary, we have demonstrated a novel and relatively simple approach to photodetection with a high photoresponsivity using a graphene phototransistor fabricated on undoped SiC substrate. The photoresponse characteristics of the device based on this new architecture show many distinct advantages, including strong and ambipolar gate voltage tunability, high photocurrent to dark-current ratio, and high photoresponsivity at room temperature. The high photoresponsivity (~7.4 A/W) of our device is not only superior to most other recently developed graphene photodetectors but also higher than the required photoresponsivity (1 A/W) in most practical applications. We anticipate that the photoresponsivity of devices based on the demonstrated approach can be further improved by optimizing the fabrication processes and measurement conditions (e.g., increasing source-drain bias voltage). In addition, our method may take advantage of a wide range of undoped semiconductors (differing in bandgaps and other electro-optical properties) as substrates for fabricating photodetectors. Our simple approach can also be generalized to other "beyond-graphene" 2D-semiconductors such as molybdenum disulfide ($MoS_2$),[41] or to higher-energy radiation.[35] Given the significant design flexibility and simplicity of our approach, this work provides a promising groundwork for the future development of graphene-based high-performance optoelectronic devices.

## Methods

**Device fabrication**: Monolayer graphene was prepared by a micromechanical exfoliation method from highly ordered pyrolytic graphite (Momentive Performance Materials Inc.) and subsequently transferred (see details of the transfer process in Supplementary Information) onto an undoped 6H (Si-faced) SiC substrate (Pam-Ximan, with typical absorption coefficient ~40/cm at wavelength of 400 nm). The source-drain contacts with channel length of ~ 2 µm and channel width of ~ 2 µm were fabricated using electron beam lithography followed by deposition of Cr (5 nm)/Au (65 nm). The back-gate contact was fabricated by deposition of Cr (5 nm)/Au (65 nm) onto the back side of SiC wafer.

**Device characterization:** The two-terminal dc transport measurements of the GFETs were performed using Keithley 2400 source meters controlled by a LabView program. The photoelectronic response was measured by illuminating the entire device by a laser with wavelength of 400 nm. The incident laser beam spot size on the device is ~ 2 mm. The laser power was tuned by controlling the laser drive current and was calibrated using a power meter.

**Acknowledgements**

The authors acknowledge partial support of this work from DHS (grant 2009-DN-077-ARI036), and DTRA (grant HDTRA1-09-1-0047). We thank Sourav Dutta for providing the laser and help with the measurement setup.

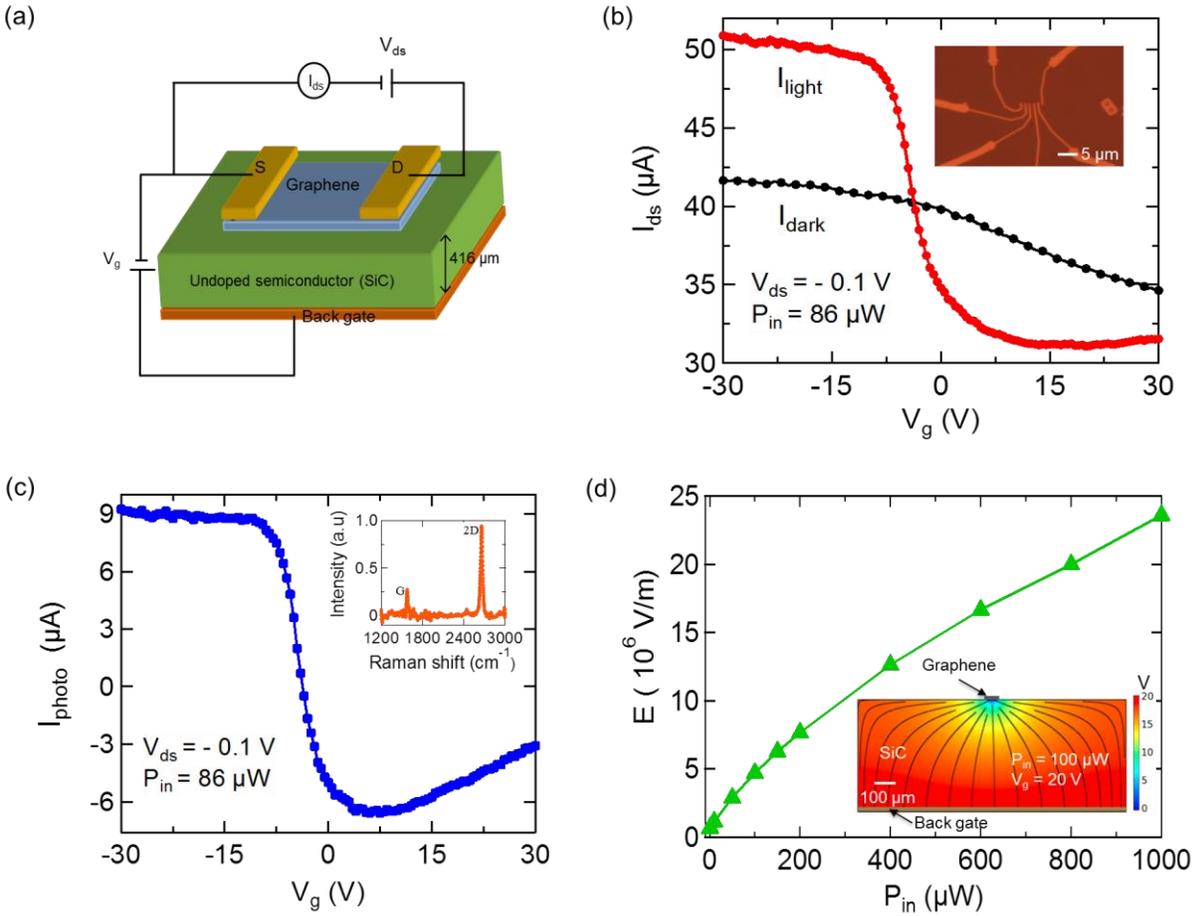

**Figure 1 | Gate voltage tunable photoresponse and operational principle of a graphene phototransistor on undoped semiconducting substrate. a.** Schematic of a graphene field effect transistor (GFET) on an undoped semiconductor substrate. In this work, an undoped silicon carbide (SiC) is used as the substrate. A back-gate voltage is applied at the back of SiC substrate to produce an electric field acting on the graphene and modulating graphene conductivity *via* field effect. **b**. Source-drain current ($I_{ds}$) as a function of back-gate voltage ($V_g$) of a GFET on SiC substrate for a fixed $V_{ds}$ = - 0.1 V, without and with illumination of a laser (wavelength $\lambda$ = 400 nm, laser power incident on device $P_{in}$ = 86 µW). Inset: Optical microscope image of a representative GFET device (top view). **c**. The dependence of the photocurrent ($I_{photo}$) of the GFET on gate voltage. The photocurrent is extracted by subtracting the dark current ($I_{dark}$) from the light current ($I_{light}$), both shown in Fig. 1b. Inset: Raman spectrum of exfoliated graphene on a SiC substrate, indicating a single layer of graphene. **d**. A plot of electric field (*E*) (simulated with COMSOL Multiphysics) under the graphene as a function of $P_{in}$, showing an increase in *E* with increasing incident laser power $P_{in}$. Inset: A representative simulation of the electric potential (color scale) and electric field lines in a GFET with a back gate voltage ($V_g$) of 20 V and $P_{in}$ = 100 µW. The SiC thickness (416 µm) in the modeled device is the same as the thickness of SiC in our experimental devices. The scale bar for the SiC is 100 µm. The graphene and back-gate electrode are not drawn to the scale. To qualitatively account for the effects of native oxide and spatially non-uniform generation of photo carriers in the substrate, we have assumed the conductivity of the top 10 nm of the SiC to be not affected by illumination. The stream lines represent the electric field lines, which direct the

photogenerated carriers toward the location directly under the graphene. The electric field shown in the main panel (d) is calculated at the location under the graphene. The strength of the electric field under graphene increases with increasing $P_{in}$. The change in the electric field is detected by the change of conductivity of graphene, allowing us to detect the light incident on the GFET.

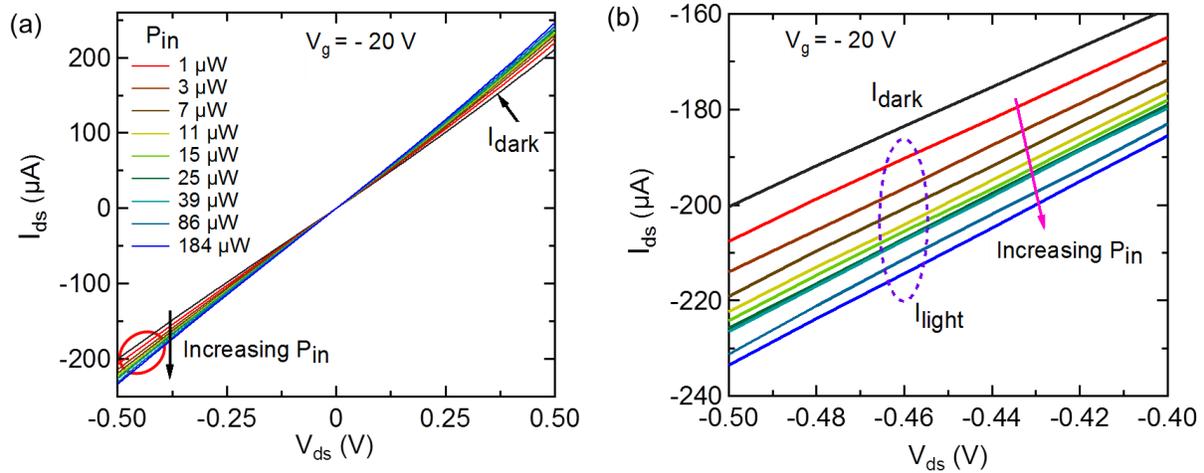

**Figure 2 | Dependence of current ($I_{ds}$) on source-drain bias voltage ($V_{ds}$) of a graphene phototransistor with increasing light power.** (a) $I_{ds}$ - $V_{ds}$ characteristics of a typical GFET at $V_g = -20$ V without and with illumination for a series of incident laser power $P_{in}$ (varying from 1 to 184 µW). All $I_{ds}$ - $V_{ds}$ curves pass through the origin, while the slope of $I_{ds}$ - $V_{ds}$ curves increase with increasing laser power. (b) Enlarged view of the circled region shown in (a), showing the increase of current ($I_{lihgt}$) under laser illumination with increasing laser power.

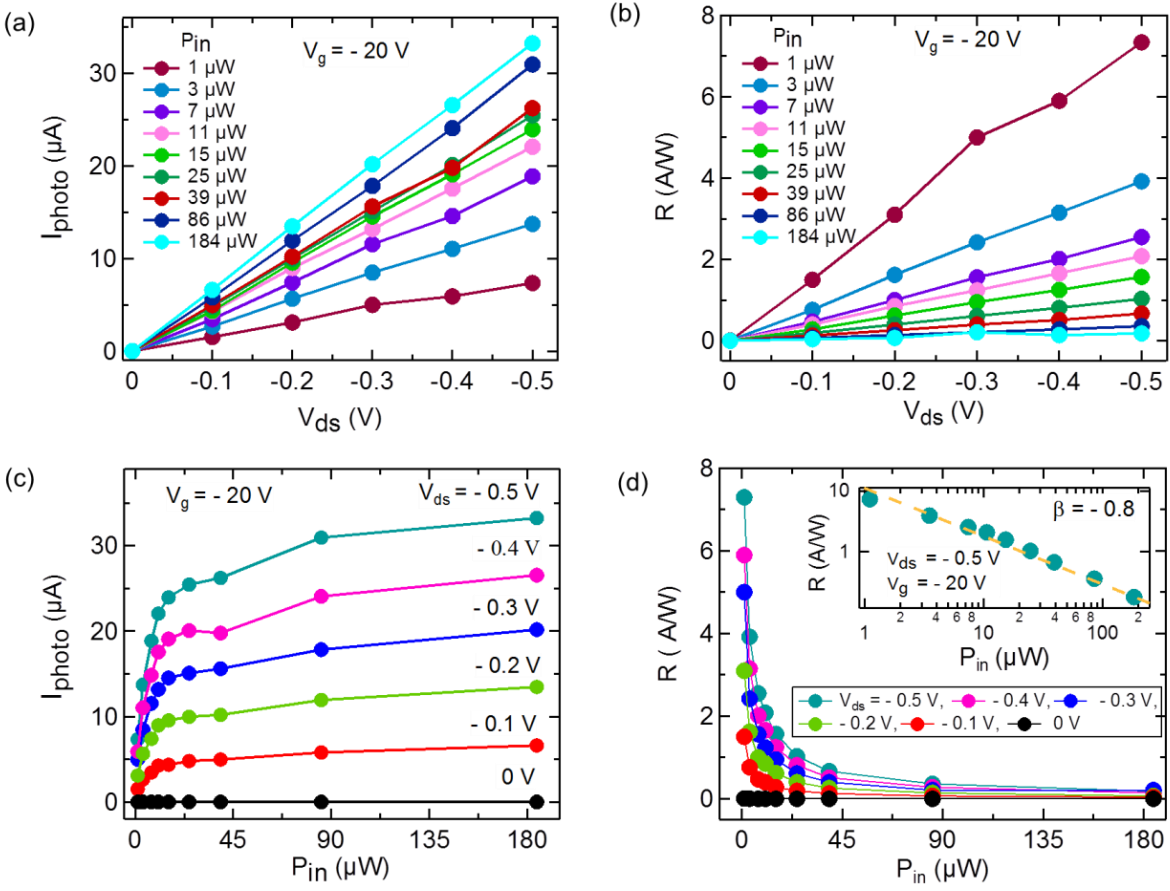

**Figure 3 | Dependence of photocurrent and photoresponsivity on source-drain bias voltage and light power.** (a) Photocurrent ($I_{photo}$) at various source-drain bias voltages $V_{ds}$ (from 0 to - 0.5 V) for a series of incident laser powers $P_{in}$ (from 1 to 184 µW) and $V_g$ = - 20 V. (b) Photoresponsivity ($R$) as a function of $V_{ds}$ for various $P_{in}$ as shown in (a). Both the photocurrent and photoresponsivity increase with increasing $V_{ds}$. A photoresponsivity of 7.4 A/W is achieved for $V_{ds}$ = - 0.5 V at $P_{in}$ = 1 µW. (c) Photocurrent as a function $P_{in}$ is shown for different $V_{ds}$ (from 0 to - 0.5 V), indicating that photocurrents saturate for higher laser powers for all $V_{ds}$. (d) The dependence of photoresponsivity on $P_{in}$ for the same $V_{ds}$ as shown in (c). Inset: A log-log plot of $R$ *versus* $P_{in}$ for $V_{ds}$ = - 0.5 V. The dashed line is a power law fit ($R \propto P_{in}^{\beta}$) to the experimental data (filled circles) with a power $\beta \sim$ - 0.8.
11

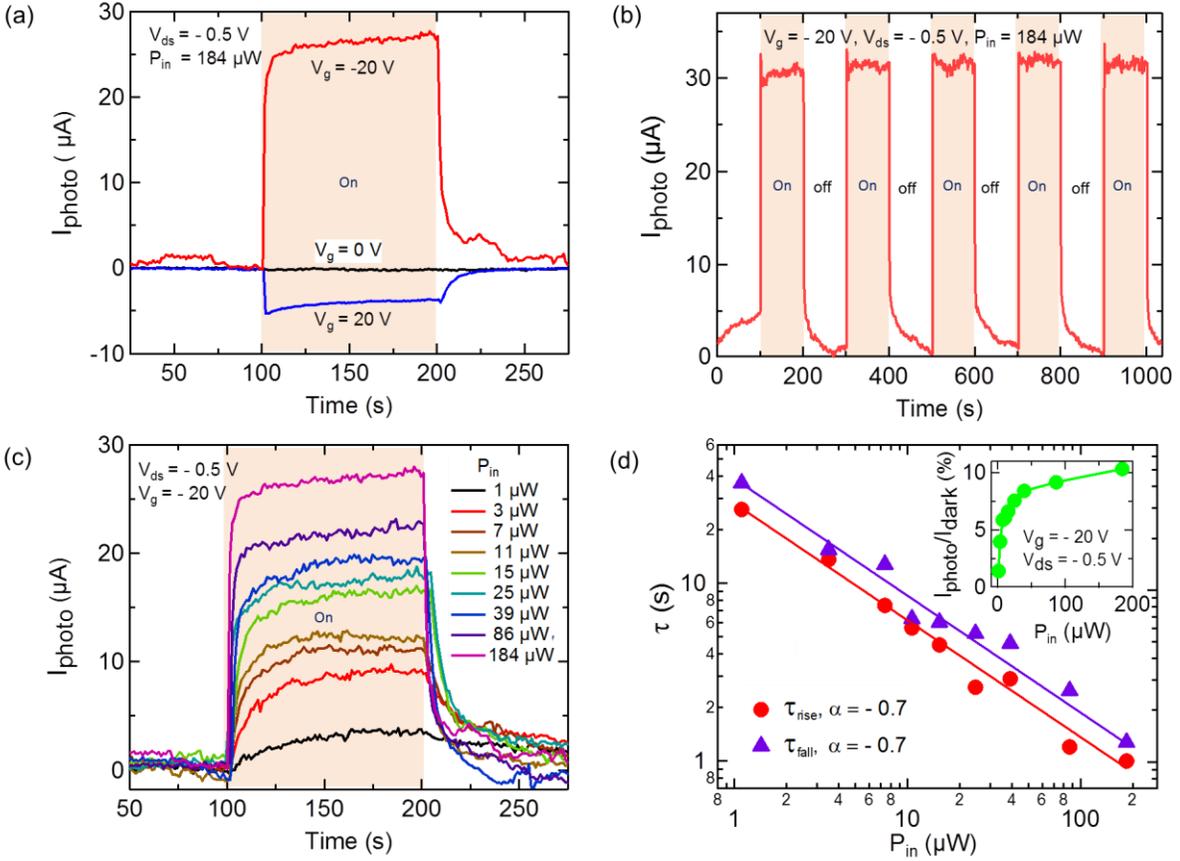

**Figure 4 | Dynamical photoresponse of a graphene phototransistor.** (a) Time-dependent photocurrent of the GFET for $V_g$ of - 20 V, 0 and 20 V, as the laser is turned on and off. A positive photocurrent is observed for $V_g$ = - 20 V, whereas a negative photocurrent is observed for $V_g$ = 20 V. Photocurrent is nearly zero for $V_g$ = 0 V. The sign of the photocurrent is consistent with the field effect measurement in Fig. 1b & c. (b) Time-dependent photocurrent as the laser ($P_{in}$ = 184 µW) is repeatedly turned on and off at $V_{ds}$ = - 0.5 V and $V_g$ = - 20 V. (c) Photocurrent as a function of time at $V_{ds}$ = - 0.5 V, $V_g$ = - 20 V and various incident laser powers $P_{in}$ (from 1 to 184 µW). Shaded regions in (a-c) mark time intervals during which the laser is on. (d) The response time ($\tau$) of the rise and fall of photocurrent dynamics is shown in (c) as a function of $P_{in}$. The shortest response time of our device is ~ 1 s. Solid straight lines represent power law fits ($\tau \propto P_{in}^\alpha$). Inset: The ratio of photocurrent to dark-current $I_{photo}/I_{dark}$ (in %) as a function of $P_{in}$. The maximum $I_{photo}/I_{dark}$ of our GFET is ~ 10.5 %, measured for laser power of 184 µW.

# Supplementary Information

# Gate-tunable and high responsivity graphene phototransistors on undoped substrates

Biddut K. Sarker,[1,2] Isaac Childres,[1,2] Edward Cazalas,[3] Igor Jovanovic,[3] and Yong P. Chen[1,2,4]

[1]Department of Physics and Astronomy, Purdue University, West Lafayette, Indiana 47907, USA
[2]Birck Nanotechnology Center, Purdue University, West Lafayette, Indiana 47907, USA
[3]Department of Mechanical and Nuclear Engineering, The Pennsylvania State University, University Park, Pennsylvania 16802, USA
[4]School of Electrical and Computer Engineering, Purdue University, West Lafayette, Indiana 47907, USA

## 1. Transfer of the exfoliated graphene onto SiC substrate

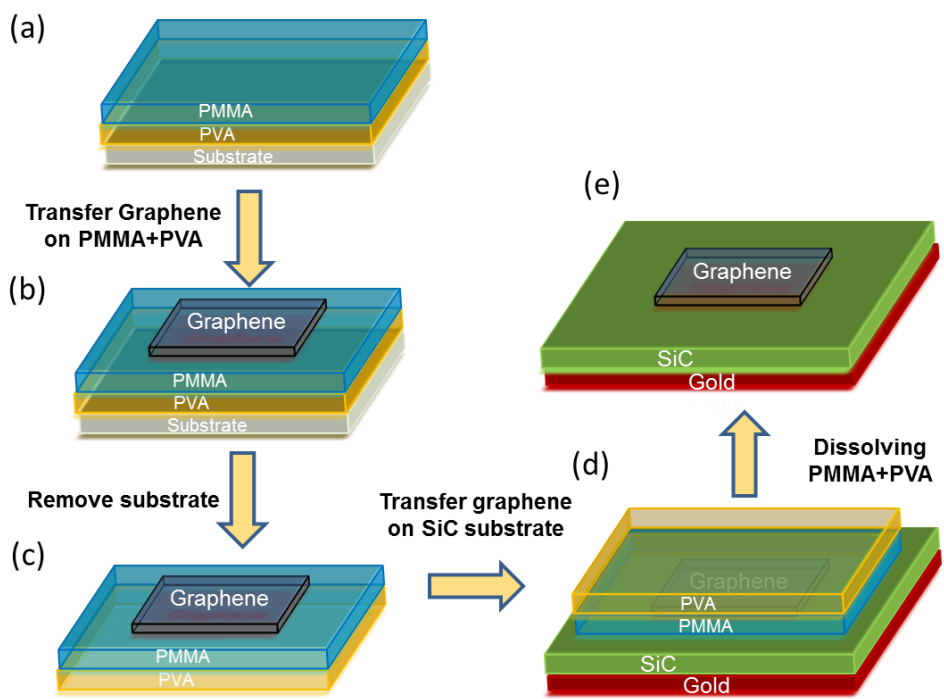

**Figure S1**: Schematic of the process to transfer an exfoliated monolayer graphene onto a SiC substrate.

We transfer an exfoliated monolayer graphene onto a silicon carbide (SiC) substrate by the following processes. First, polyvinyl alcohol (PVA) solution is coated on a sacrificial substrate (here, Si/SiO$_2$ with dimension of 2 × 2 cm is used) at 3000 rpm for 45 s and baked on a hotplate at 90 $^0$C for 5 min. Then PMMA (polymethyl methacrylate) is coated onto the PVA film and similarly baked (Fig. S1a). Monolayer graphene was prepared by a micromechanical exfoliation technique and transferred onto the polymer (PMMA + PVA) films (Fig. S1b). The polymer films containing the graphene are then separated from the sacrificial substrate (Fig. S1 c) and transferred onto an undoped SiC substrate using a homemade transfer stage (Fig. S1d). Finally, the SiC substrate is submerged in acetone for a few hours to remove the polymer films (PMMA + PVA), then rinsed with IPA (isopropyl alcohol) and blown dry with nitrogen gas (Fig. S1e).

## 2. Raman characterization of the graphene on SiC substrate

We used Raman spectroscopy to confirm that the transferred exfoliated graphene on the SiC substrate is a monolayer. The Raman spectrum is measured using a Horiba Jobin Yvon Xplora confocal Raman microscope with a 532 nm excitation laser. Spectra were taken under the same experimental conditions on the same device at two different spots; one spot is on the graphene on the SiC substrate, and the other spot is on the SiC substrate (where no graphene is present) (inset in Fig. S2b). Since the intensity of Raman peaks varies slightly from spot to spot, the spectra were normalized by the strongest peaks. Since the Raman spectra of graphene and SiC have substantial overlap with each other, we subtracted the normalized SiC spectrum (Fig. S2a) from the normalized spectrum of graphene on the SiC (graphene + SiC) (Fig. S2b).

The difference of the normalized spectrum of SiC, and graphene on the SiC is the graphene spectrum, which is shown in the inset of Fig. 1c in the main manuscript. The graphene spectrum shows no D peak, suggesting negligible defects in graphene.[1] The ratio of the 2D to G peaks intensity ($I_{2D}/I_G$) of the graphene spectrum is more than two, indicating a monolayer graphene in our device (inset, Fig. 1c). [1, 2]

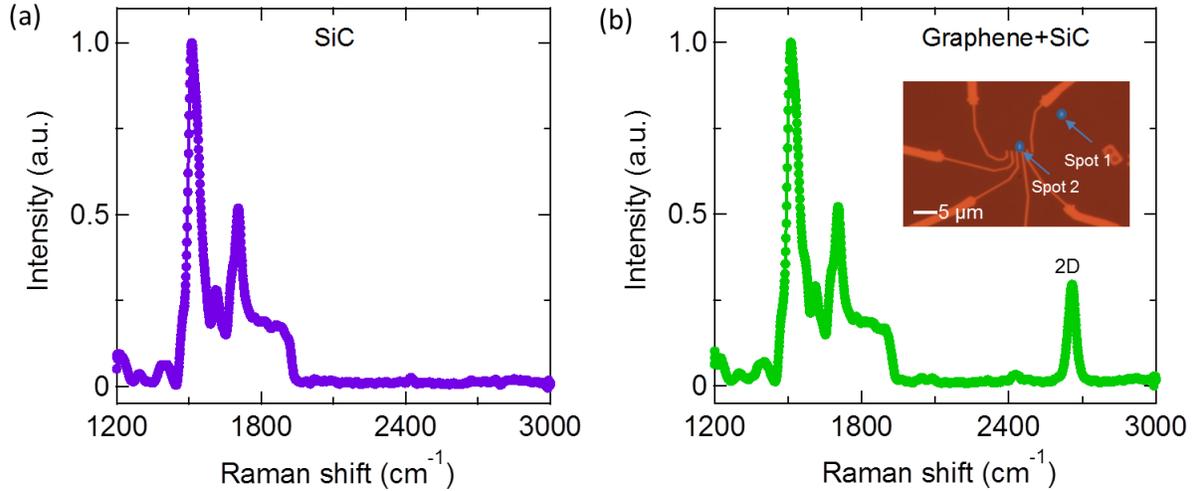

**Figure S2**: Raman spectrum of (a) SiC substrate (without graphene) and (b) graphene on SiC. Inset of (b): Optical image of a fabricated device. Spot 1 (SiC) and 2 (graphene on SiC) show where the Raman spectra were taken. The spectrum intensity is normalized by its strongest peak. The difference between the spectra (b) and (a) is extracted as the graphene Raman spectrum and shown in the inset of Fig. 1c.

## 3. Calculation of Conductivity

The conductivity of the SiC substrate increases by absorption of the incident light, whereby electrons and holes are produced in the SiC. The change in SiC conductivity due to light illumination can be calculated by $\Delta\sigma = q\mu = q'\tau\mu$, where $q(=q'\tau)$ is the number of steady state carriers (for both the electrons and holes) produced per unit volume and $q'$ is the number of carriers produced per unit volume per unit time through light absorption, $\mu$ (= 400 cm$^2$/V.s, given by manufacturer, PAM-Xiamen) is the mobility within SiC and $\tau$ is the carrier life time (recombination time), the mean time a conductive charge may exist within the substrate before recombination with an opposite charge.

Here we assume that $\tau = 1$ μs [ref. 3,4]. We consider the influence of penetration depth of the light in the SiC substrate. For simplicity, we divided the total thickness of SiC substrate into three parts and the profile of light absorption throughout the depth of the SiC substrate is used to calculate charge density produced per unit time for each part. For example, the time dependent number of change carrier per unit volume for the top 1/3 of SiC substrate is calculated to be $q' = 1.04 \times 10^3$ C/m$^3$/s for $P_{in} = 1$ µW. The final SiC conductivity of the top 1/3 SiC substrate due to laser irradiation is given as $\sigma = \sigma_t + \Delta\sigma = 1\times10^{-3} + P_{in}(4.2\times10^{-5})$ S/m, where $P_{in}$ is in µW and $\sigma_t$ is typical value of un-irradiated SiC conductivity, $\sigma_t \approx 1\times10^{-3}$ S/m is given the manufacturer (PAM-Xiamen).

## 4. Gate leakage current of device in dark and under laser illumination

To confirm that gate leakage current ($I_g$) is not contributing to the photocurrent of our device, $I_g$ is monitored in both the dark and light illumination conditions. Figure S3a, b and c show the plots of $I_g$ as a function of gate-voltage ($V_g$), $I_g$ as a function of source-drain bias ($V_{ds}$) and $I_g$ as a function of time, respectively, measured in the dark and under illumination with various incident laser powers on the device ($P_{in}$).

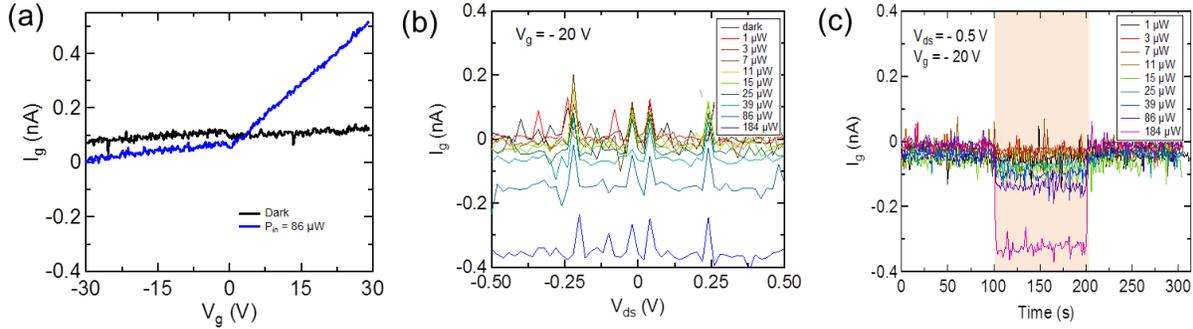

**Figure S3**: (a) Gate leakage current ($I_g$) *vs.* gate voltage ($V_g$), (b) $I_g$ *vs.* source-drain bias voltage ($V_{ds}$), (c) $I_g$ *vs.* time, measured in the dark and under illumination with various incident laser powers ($P_{in}$) ranging from 1 to 184 µW. The shaded region in c labels the time interval when the laser is turned on.

From these plots (Fig. S3a-c), we found that the device (i) leakage current is small (<1 nA) compared to the measured photocurrent (in the range of µA), and that (ii) leakage currents both in the dark and under laser illumination with a low laser power are almost similar. These features indicate that leakage current does not increase significantly with a low incident laser power. While leakage current does increase for a higher laser powers, it remains less than 1 nA. We therefore conclude that gate leakage current does not contribute to the measured photocurrent in our device.

## 5. Control experiment 1: SiC device (without graphene)

To confirm the photoresponse of our GFETs is not due to the photoresponse of the substrate (SiC) or Schottky contact at the SiC/metal interface, we fabricated SiC control devices without graphene (making a direct contact on top of SiC). Optical image of a fabricated SiC device (without graphene) is shown in the inset of Fig. S4a. The plots of $I_{ds}$ - $V_g$ and $I_{photo}$-$V_g$ characteristics of a representative SiC device with and without illumination are shown in Figure S4a and b, respectively. We use same scale in the Fig.S4b and Fig. 1c ($I_{photo}$-$V_g$ of GFET in the main figure) in order to clearly show the difference between the photocurrents and their gate dependence in the GFET and SiC devices.

These plots show that both the current (in dark as well as under illumination) and photocurrent of SiC device are very small (of the order nA), at least three order lower than that observed photocurrent in the GFET (of the order µA). In addition, the photocurrent in the SiC device does not change significantly with the gate voltage, whereas the photocurrent in GFETs shows a strong gate-voltage dependence (Fig. 1c). From those measurements we can conclude that the photoresponse of our GFETs does not result from the Schottky contact of SiC or the collection of the charge from the SiC; rather it is the result of the modulation of charge carriers in the graphene *via* field effect.

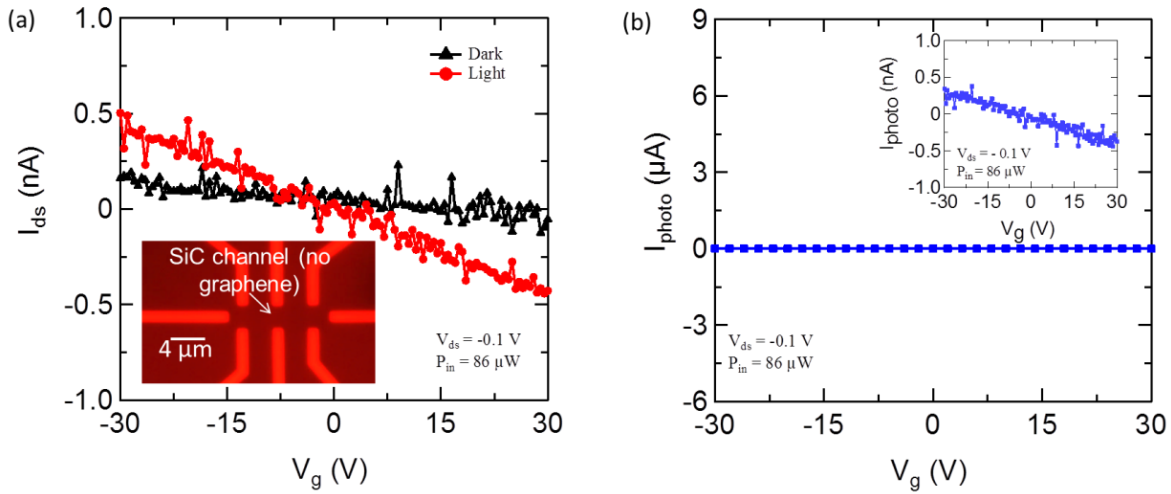

**Figure S4.** (a) Dependence of the drain-source current on the back-gate voltage ($I_{ds}$ - $V_g$) of a SiC device (without graphene) without and with laser illumination ($\lambda$ = 400 nm) at incident laser power ($P_{in}$) of 86 µW. Inset: Optical image of a fabricated SiC device (without graphene). (b) The dependence of the photocurrent ($I_{photo}$) of the GFET on the gate voltage. The scales of Fig.4Sb and Fig. 1c (in the main paper) are kept the same to clearly show the differences in the gate dependence of photocurrent generation. Inset: the same Fig.4Sb plot, but on a nA scale.

## 6. Control experiment 2: Dummy device (gold in the channel instead of graphene)

We also fabricated dummy devices that contain no graphene, but use gold as a channel between the source-drain contacts (Inset of Fig. S5a). The $I_{ds}$ - $V_g$ plot of the dummy device shows no change in source-drain currents with the gate-voltage, with and without illumination. The $I_{ds}$ - $t$ characteristics also show no change in current under light illumination (Fig. S5b). These observations further confirm that graphene is essential for the field effect photoresponse observed in the GFETs, and the photoresponse does not come from the SiC substrate or SiC/Au interface.

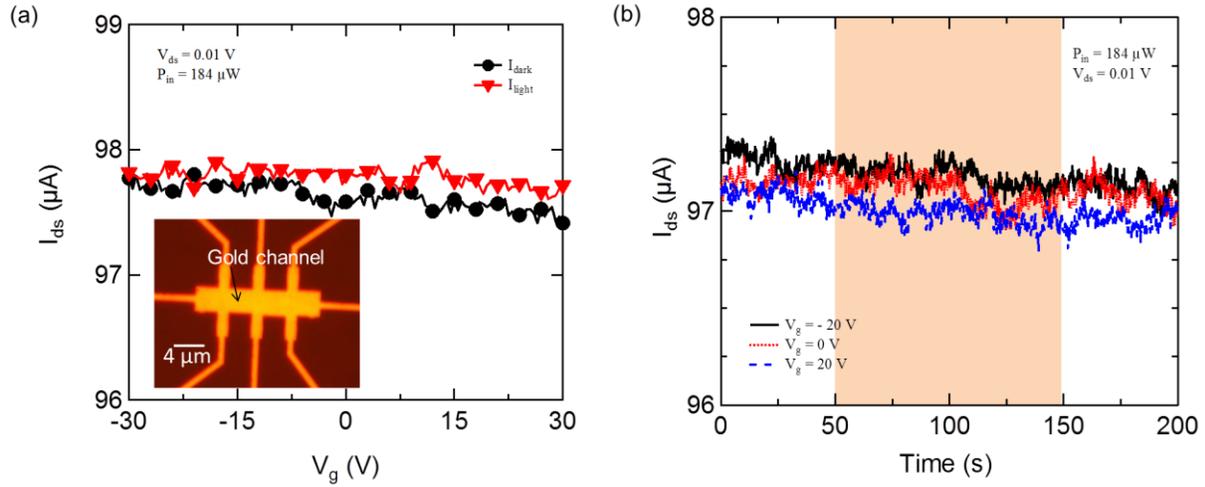

**Figure S5.** (a) $I_{ds}$ - $V_g$ characteristics of a dummy device without and with laser illumination ($\lambda$ = 400 nm). Inset: Optical image of a fabricated dummy device (gold in the channel instead of graphene) (b) $I_{ds}$ - $t$ characteristics for different the gate voltages when the laser switches on and off. Shaded area indicates the time intervals during which the laser is on.

## 7. Modulated charge carriers per incident photon

The number of modulated charge carriers in the device per incident photon is calculated using the formula,[5] $(I_{photo}/P_{in}) \times (hc/e\lambda)$; where, $I_{photo}$ is the photocurrent, $P_{in}$ is the incident laser power on the device, $h$ is Planck's constant, $e$ is electron charge, and $\lambda$ is the wavelength of incident light (400 nm). The number of modulated charge carriers per incident photon increases with increasing source-drain bias voltage because photocurrent increases with increasing the source-drain bias. We found that for a source-drain bias of -0.5 V, approximately 23 electrons (or holes) can be modulated in graphene by a single photon incident into the SiC substrate of our device.

## 8. Photocurrent response time

The photocurrent ($I_{photo}$) rise and fall response times ($\tau$) for all laser powers was calculated by fitting the photocurrent *vs.* time data to an exponential function. Four representative fitted curves (for both rise and fall) for two different laser powers of 184 µW and 25 µW are shown in Fig. S6. The rise and fall times for $P_{in}$ = 184 µW are found to be 1.0 and 1.3 s, respectively, whereas for $P_{in}$ = 25 µW, the rise and fall time are 2.6 and 5.6 s, respectively.

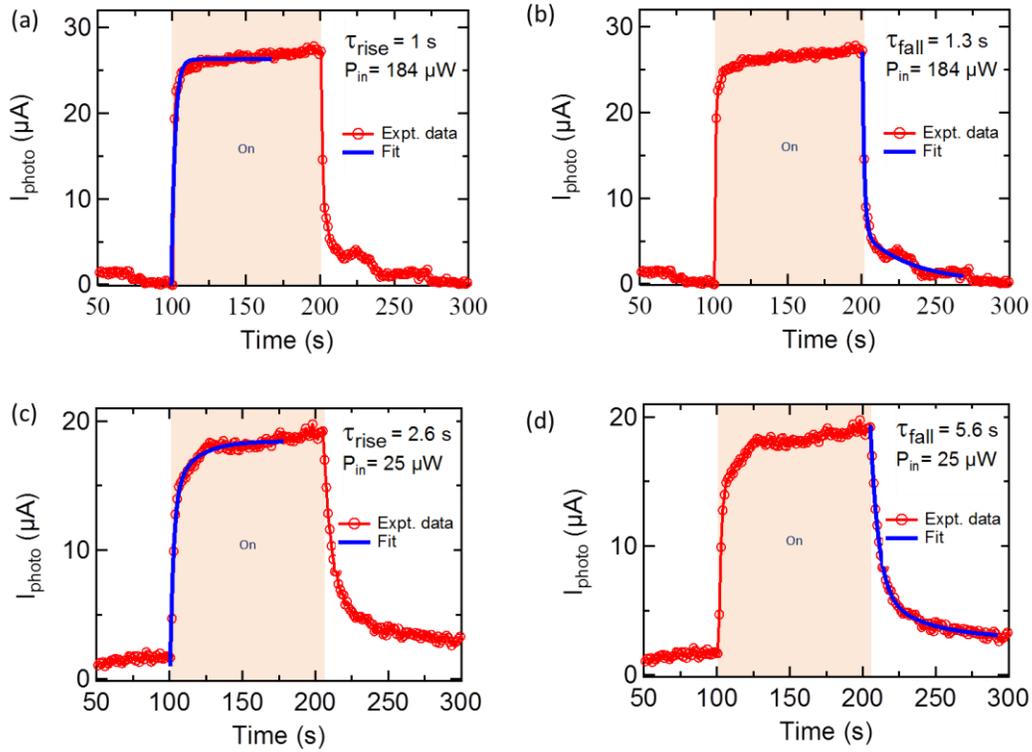

**Figure S6**: (a-d) Photocurrent ($I_{photo}$) *vs.* time for $V_g = -20$ V and $V_{ds} = -0.5$ V (a, b) for incident laser powers ($P_{in}$) of 184 μW and (c, d) $P_{in} = 25$ μW. Red circles are experimental data; solid lines represent exponential fits to extract the time constants. Shaded regions in a-d label time intervals during which the laser is on.